\documentclass{webofc}
\usepackage{hyperref}
\usepackage[varg]{txfonts}   % Web of Conferences fontt particle symbols

\usepackage{xspace} 
\usepackage{upgreek}
\usepackage{ifthen}
\newboolean{uprightparticles}
\setboolean{uprightparticles}{false}

 \def\thebaroffset{0.0em}
\newcommand{\offsetoverline}[2][\thebaroffset]{\kern #1\overline{\kern -#1 #2}}%
\def\CP                {{\ensuremath{C\!P}}\xspace}            
\def\PB      {\ensuremath{B}\xspace}
\def\PD      {\ensuremath{\mathrm{D}}\xspace} 
\def\Pd      {\ensuremath{d}\xspace}
\def\Ps      {\ensuremath{s}\xspace}  
\def\Pb      {\ensuremath{\mathrm{b}}\xspace}
\def\PJ      {\ensuremath{J}\xspace} 
\def\PK      {\ensuremath{\mathrm{K}}\xspace}
\def\Ppsi        {\ensuremath{\uppsi}\xspace} 
\def\Ppi         {\ensuremath{\pi}\xspace} 
\def\dquark    {{\ensuremath{\Pd}}\xspace}
\def\dquarkbar {{\ensuremath{\overline \dquark}}\xspace}
\def\ddbar     {{\ensuremath{\dquark\dquarkbar}}\xspace}
\def\squark    {{\ensuremath{\Ps}}\xspace}
\def\squarkbar {{\ensuremath{\overline \squark}}\xspace}
\def\ssbar     {{\ensuremath{\squark\squarkbar}}\xspace}
\def\bquark    {{\ensuremath{\Pb}}\xspace}
\def\bquarkbar {{\ensuremath{\overline \bquark}}\xspace}
\def\bbbar     {{\ensuremath{\bquark\bquarkbar}}\xspace}
\def\B       {{\ensuremath{\PB}}\xspace}
\def\Bbar    {{\ensuremath{\offsetoverline{\PB}}}\xspace}

\def\BorBbar {\kern \thebaroffset\optbar{\kern -\thebaroffset \PB}\xspace}
\def\Bz      {{\ensuremath{\B^0}}\xspace}

\def\Bu      {{\ensuremath{\B^+}}\xspace}

\def\Bp      {{\ensuremath{\Bu}}\xspace}
\def\Bpm     {{\ensuremath{\B^\pm}}\xspace}
\def\Bs      {{\ensuremath{\B^0_\squark}}\xspace}
\def\Bsb     {{\ensuremath{\Bbar{}^0_\squark}}\xspace}

\def\jpsi     {{\ensuremath{{\PJ\mskip -3mu/\mskip -2mu\Ppsi\mskip 2mu}}}\xspace}
\def\kaon    {{\ensuremath{\PK}}\xspace}
\def\Kstarz  {{\ensuremath{\kaon^{*0}}}\xspace}
\def\Kp      {{\ensuremath{\kaon^+}}\xspace}

\def\pion   {{\ensuremath{\Ppi}}\xspace}

\def\pim    {{\ensuremath{\pion^-}}\xspace}

\def\Ds      {{\ensuremath{\D^+_\squark}}\xspace}
\def\D       {{\ensuremath{\PD}}\xspace}

\def\lhcb   {\mbox{LHCb}\xspace}

\def\lhc    {\mbox{LHC}\xspace}
\def\bfactories {\mbox{\B Factories}\xspace}

\begin{document}
\title{Fast Inclusive Flavour Tagging at \lhcb}
%
% subtitle is optionnal
%
%%%\subtitle{Do you have a subtitle?\\ If so, write it here}

\author{\firstname{Claire} \lastname{Prouve}\inst{1}\fnsep\thanks{\email{claire.prouve@cern.ch}} \and
        \firstname{Niklas} \lastname{Nolte}\inst{2} \and
        \firstname{Christoph} \lastname{Hasse}\inst{3}
        % etc.
}

\institute{Instituto Galego de F\'isica de Altas Enerx\'ias (IGFAE), Universidade de Santiago de Compostela, Santiago de Compostela, Spain
\and Massachusetts Institute of Technology, Cambridge, MA, United States
\and European Organization for Nuclear Research (CERN), Geneva, Switzerland}
\abstract{%
The task of identifying \B meson flavour at the primary interaction point in the \lhcb detector is crucial for measurements of mixing and time-dependent \CP violation. Flavour tagging is usually done with a small number of expert systems that find important tracks to infer the \B meson flavour from. Recent advances show that replacing all of those expert systems with one ML algorithm that considers all tracks in an event yields an increase in tagging power. However, training the current classifier takes a long time and is not suitable for use in real-time triggers. In this work we present a new classifier, based on the DeepSet architecture. With the right inductive bias of permutation invariance, we achieve great speedups in training (multiple hours vs 10 minutes), a factor of 4-5 speed-up in inference for use in real time environments like the trigger and less tagging asymmetry. For the first time we investigate and compare performances of these “Inclusive Flavor Taggers” on simulation of the upgraded \lhcb detector for the third run of the \lhc.

}
\maketitle

\section{Introduction}
\label{sec:introduction}
The identification of the flavour of \Bz and \Bs mesons at production is crucial for many \B-mixing and time-dependent \CP-violation measurements~\cite{Bsoscil} ~\cite{Bs2JpsiK}. The identification of the \B meson flavour at production relies on analysing information from the rest of the event. The procedure of determining the flavour of a \B meson at the time of its production utilising information from the rest of the event is called \textit{flavour tagging}.
\\
\\
At the \bfactories, flavour tagging is done with high efficiency since the vast majority of \B mesons are produced as quantum-correlated pairs via the decay of $\Upsilon(4S)$ or $\Upsilon(5S)$ resonances. If the flavour of one \B meson is identified, then the other can be inferred. At proton-proton colliders this is more difficult as not all \B mesons are produced in \B \Bbar pairs, and of those produced in pairs, most are not produced in quantum-correlation. Additionally, unlike at the \bfactories, the reconstruction of the second \B meson - if it exists - cannot be performed with high efficiency. Combined with the necessarily higher background from uninformative `background' particles stemming from the proton-proton collision, this makes flavour tagging at \lhcb considerably more difficult.
\\
\\
In these proceedings we lay out the general way flavour tagging works at \lhcb as well as the specific algorithms used in the past. Then we present the new approach of the inclusive tagger in the specific implementation using a DeepSet Neutral Network as well as its advantages and compare its performances to past taggers. We end with a summary and conclusion.

\section{Flavour Tagging Information in the Event}
\label{sec:ftathadron}
\begin{figure}[h]
% Use the relevant command for your figure-insertion program
% to insert the figure file.
\centering
\includegraphics[width=\textwidth, trim={0 7.3cm 13cm 0cm},clip]{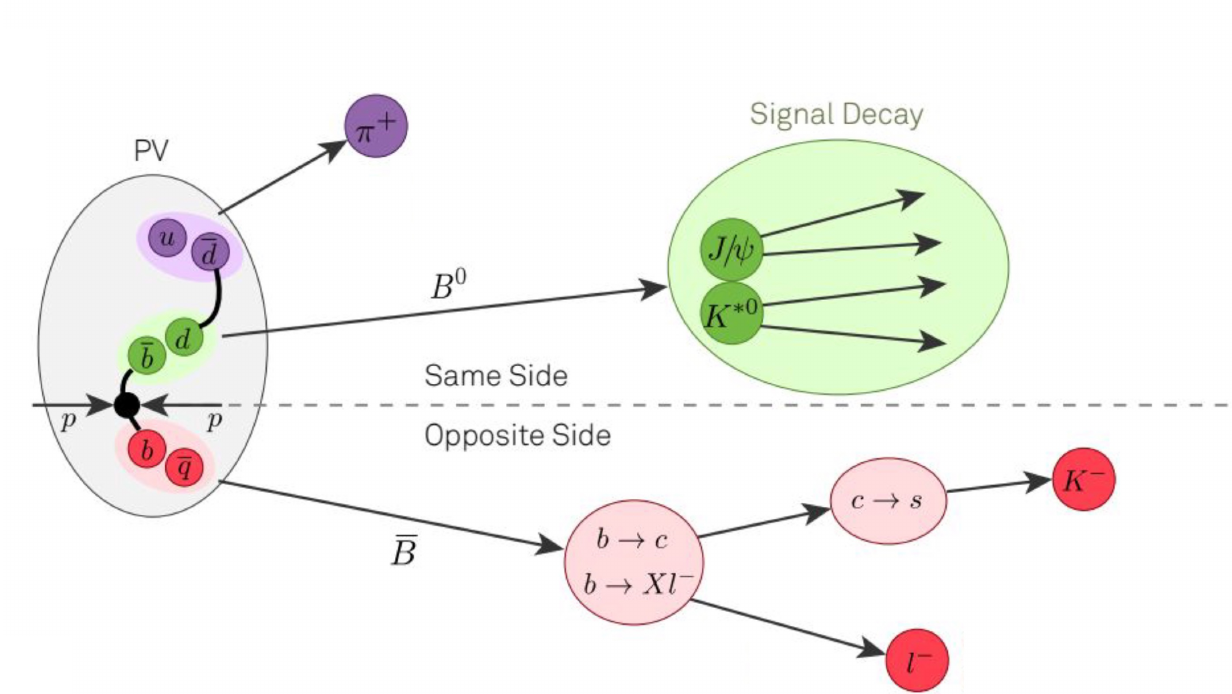}
\caption{Illustration of the flavour tagging information present in the event.}
\label{fig:fig1}       % Give a unique label
\end{figure}

Information on the flavour of the signal \B meson can be present in different ways in the event. This is illustrated in Figure~\ref{fig:fig1}. The signal \B meson decay is pictured on the top half of the illustration. The top half is therefore referred to as the \textit{same side}. The bottom half is called the \textit{opposite side} and contains another \B meson decay, referred to as the \textit{opposite-side \B meson}.

Even in proton-proton collision \bquark quarks are usually produced in \bbbar pairs. At \lhcb 24\% of events have both, \bquark and \bquarkbar quarks, produced within the detector acceptance. The \bquark and \bquarkbar quarks hadronise to produce two \B hadrons -- the signal \B and the opposite-side \B \ -- of opposite flavour. By determining the flavour of the opposite-side \B meson, the flavour of the signal \B meson can be inferred as being the opposite. This strategy is also employed by the \bfactories.

In addition to the opposite side information, same side information is present uniquely in environments in which \B mesons are produced via the strong interaction, such as proton-proton colliders, but not the \bfactories. In the hadronisation process of the signal \bquark quark, additional particles are produced which are correlated in phase-space with the signal decay itself. These particles are called \textit{same-side} tagging particles. If the same-side particles can be reconstructed and identified the flavour of the signal \B meson can be inferred from the charge of the same-side \textit{tagging particle}. For \Bz mesons, the same-side tagging particle is a pion, formed from the \ddbar quark pair from the hadronisation process. The \dquark quark is used with the \bquarkbar to form the \Bz meson, and the left over \dquarkbar forms a positively charged pion (equivalently a \Bbar meson is produced together with a negatively charged pion). Similarly, for \Bs mesons, the light meson associated with the hadronisation is a positively charged kaon, produced with the \squark quark from the \ssbar quark pair from the hadronisation process (and a negatively charged kaon for the \Bsb). About 50\% of \Bz mesons are accompanied by a charged pion and 50\% of \Bs mesons by a charged kaon. 
%\footnote{Additionally, a pion can be formed from the decay of a $B^*$ state, but the signature is the same.}

\section{Classical Taggers}
\label{sec:classicaltaggers}
The traditional approach to flavour tagging at \lhcb is what we call the \textit{classical taggers}~\cite{sstaggers}~\cite{ostaggers}. Each of the classical taggers is an algorithm that attempts to identify \textit{one} specific tagging particle in the event that carries information about the flavour of the signal \B meson. The same side taggers search for charged particles that were produced during the hadronisation process of the signal \B meson. These charged particles are kinematically correlated with the signal decay. The opposite side taggers search for specific decay products from the opposite-side \B meson. 
All classical taggers perform a selection on all charged particles in the event. Then a multivariate analysis tool (usually a Boosted Decision Tree) is used to determine the probability that the selected particle yields the correct tagging decision. 

For each signal \B meson all classical taggers that apply for that \B meson species are run. Table~\ref{tab:classtag} shows a list of all different classical taggers and the \B mesons species they can be used for\footnote{Charged \Bpm mesons are self-tagging and don't require flavour tagging algorithms. They can therefore be used to train, validate and calibrate the taggers.}. If several taggers yield a tagging decision for the same signal \B meson, the tagging decision of the tagger with the smallest predicted probability of being wrong is chosen.
One disadvantage of the classical tagger is that each taggers aims at identifying only one specific particle in the event. That particle might not be found -- either because it was not created in the first place or because it was not produced within the detector acceptance, or because it could not be reconstructed and identified by the corresponding algorithm. As is shown in Section~\ref{sub:taggingperformance}, even when combining all taggers, no tagging decision can be reached for a significant fraction of the events. Additionally the classical taggers require the training, validation and calibration of an entire list of individual taggers.

\begin{table}
\caption{List of classical taggers and the \B meson species they can be used for.}
\centering
\label{tab:classtag}  
\begin{tabular}{l|c}
Opposite side kaon tagger & \Bz, \Bpm, \Bs \\
Opposite side muon tagger & \Bz, \Bpm, \Bs \\
Opposite side electron tagger & \Bz, \Bpm, \Bs \\
Same side kaon tagger & \Bs\\
Same side pion tagger & \Bz, \Bpm\\
Same side proton tagger & \Bz \\
\end{tabular}
\end{table}

\section{DeepSet Neutral Network Inclusive Tagger}
\label{sec:deepsetnn}
In order to address the disadvantages of the classical taggers the concept of the \textit{inclusive tagger} is introduced. The inclusive tagger considers all particles in the event simultaneously and has therefore an increased probability of reaching a tagging decision with respect to the classical taggers. Since the number of additional particles in the event varies from event to event the inclusive tagger has to be able to take a variable number of input particles. Additionally, the tagging decision should not depend on the order of the inputs, therefore the inclusive tagging algorithm has to be invariant under the permutation of the inputs. Lastly, the inclusive tagging algorithm should be fast to train and to evaluate since in the future we will want to use it in the real-time environment of the \lhcb software trigger.

\begin{figure}[h]
% Use the relevant command for your figure-insertion program
% to insert the figure file.
\centering
\includegraphics[width=\textwidth]{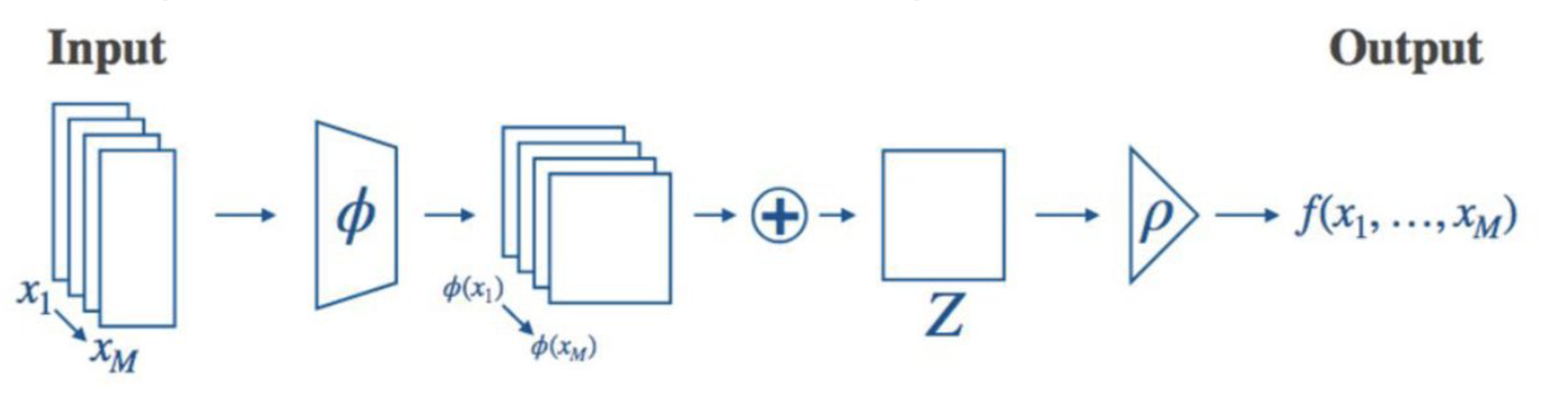}
\caption{Illustration of the DeepSet Neutral Network (DeepSet\,NN). The DeepSet\,NN acts on a list of inputs $(x_1, ... x_M)$ where $M$ can vary between different events, and $\phi$ and $\rho$ are neural networks.}
\label{fig:deepnn}       % Give a unique label
\end{figure}

An algorithm that fulfills these requirements is the DeepSet Neutral Network (DeepSet\,NN). The functionality of the DeepSet\,NN is illustrated in Figure~\ref{fig:deepnn} and presented in rigorous detail in Reference~\cite{RefDeepSetNN}. In a first step, the representation $x_i$ of each charged particle $i$ in the event\footnote{Excluding charged particles that belong to the signal \B meson decay.} is transformed individually by a neutral net $\phi$ into some representation $\phi(x_i)$. The representations $\phi(x_i)$ are then summed up. The sum is processed by another network $\rho$ to give the output of the DeepSet\,NN $f(x_1, ..., x_M)$. Therefore the structure of the DeepSet\,NN can be expressed as 
\begin{equation}
   f(x_1, ..., x_M) = \rho(\sum_i^M \phi(x_i))
\end{equation}
for an event with $M$ charged particles that do not belong to the signal \B meson decay, where $x_i$ is the representation of charged particle $i$ and $\phi$ and $\rho$ are neutral networks.

The construction of the DeepSet\,NN has one component for each input particle individually (in the form of $\phi$) and one component that acts on the event as a whole (in the form of $\rho$). Due to the summing of the $\phi(x_i)$ the DeepSet output is invariant under the permutation of the input particles. Additionally, the architecture of the DeepSet\,NN makes it easy to parallelise the training and the evaluation. Notably, it takes about an hour to train on a statically significant sample and 7$\mu s$ to evaluate per event\footnote{Previous implementations of the inclusive tagger using a different architecture took several days to train and 100$\mu s$ per event to evaluate.}.

\section{Flavour Tagging Performance}
\label{sec:ftperformance}
In this section the performance of the DeepSet\,NN tagger is compared to the performance of the classical taggers. First the performance metrics used in flavour tagging is introduced and the data-samples used for training and evaluation are presented. Then a summary of the performance numbers is given.

\subsection{Performance Metrics}
\label{sub:performancemetrics}
The tagging performance is evaluated in terms of three parameters, namely the \textit{tagging efficiency}, the \textit{mistag rate} and the \textit{tagging power}.

The tagging efficiency $\epsilon_{tag}$ is the fraction of events for which a tagging decision can be reached\footnote{Note that the tagging efficiency only encapsulates that the taggers have produced a decision, not if the decision is correct.}. This quantity is especially important for the classical taggers since the tagging particle for a specific tagger might not be present or identifiable in the event, e.g. not all events have the opposite side \B meson in the detector acceptance. The tagging efficiency is defined as 
\begin{equation}
    \epsilon_\mathrm{tag} = \frac{N_\mathrm{tagged}}{N_\mathrm{tagged} + N_\mathrm{untagged}}
\end{equation}
where $N_\mathrm{tagged}$ is the number of signal events where a tagging decision is reached and $N_\mathrm{untagged}$ is the number of signal events for which a tagging decision can not be reached. 

The mistag rate $\omega$ is the fraction of tagged events for which the tagging decision is wrong. The mistag rate is calculated as
\begin{equation}
    \omega = \frac{N_\mathrm{tagged}^\mathrm{incorrect}}{N_\mathrm{tagged}} = \frac{N_\mathrm{tagged}^\mathrm{incorrect}}{N_\mathrm{tagged}^\mathrm{correct} + N_\mathrm{tagged}^\mathrm{incorrect}}
\end{equation}
where $N_\mathrm{tagged}^\mathrm{incorrect}$ is the number of signal events where the tagging decision reached by the tagger is incorrect and $N_\mathrm{tagged}^\mathrm{correct}$ is the number of signal events where the tagging decision is correct.

The tagging power $\epsilon_\mathrm{eff}$ combines the tagging efficiency and the mistag rate into a quantity that represents the effective power of the signal sample after tagging. The tagging power is defined as
\begin{equation}
    \epsilon_\mathrm{eff} = \epsilon_\mathrm{tag} \cdot (1- 2 \cdot\omega)^2 \quad .
\end{equation}
Due to the imperfect tagging efficiency and mistag rate -- i.e. the lack of knowledge of the true flavour of the signal \B meson at production -- the statistical power of a sample of $N$ events is reduced to $\epsilon_\mathrm{eff} \cdot N$. This in turn affects measurements of e.g. \CP-violating quantities whose uncertainties scale like $1 / \sqrt{\epsilon_\mathrm{eff} \cdot N}$. Therefore, the larger the tagging power, the more precise the measurement.

Typically, opposite-side tagging algorithms have a low tagging efficiency, as these require that the opposite-side \B meson (and its decay products) are present and reconstructible in the event and identified by the algorithm; but also have a low mistag rate, as once these requirements are met, identification of the signal \B meson flavour is comparatively easy. Conversely, a pion (kaon) track close to the \Bz (\Bs) meson signal vertex can be identified in most events, however the conversion of this to a correct tag is more difficult, and therefore the same-side tagging algorithms have a generally high tagging efficiency but also a high mistag rate.

\subsection{Training and Evaluation Datasets}
\label{sub:trainingandeval}
In our study all taggers are trained and evaluated on simulated data that represents the data-taking conditions of Run\,2 (2015 - 2018) and Run\,3 (2022 - 2025) of the \lhc and the \lhcb experiment. During Run\,2, the \lhcb experiment collected data from proton-proton collisions at a fixed pileup\footnote{Pileup is the average number of proton-proton interactions per collision.} of $\sim$1. For Run\,3, the \lhcb experiment underwent an upgrade where many parts of the detector were replaced to meet the requirements of running at a higher instantaneous luminosity and at a pileup of $\sim$6.

Different signal decays are simulated and used for testing and training. The decays\footnote{Charge conjugation is implied throughout.} are $\Bz \rightarrow \jpsi \Kstarz$, $\Bp \rightarrow \jpsi \Kp$ and $\Bs \rightarrow \Ds \pim$ for Run\,2 and $\Bp \rightarrow \jpsi \Kp$ for Run\,3.

\subsection{Flavour Tagging Performance}
\label{sub:taggingperformance}
The comparison of tagging efficiency and tagging power between the classical taggers and the DeepSet\,NN tagger are shown in Tables~\ref{tab:run2perf} and~\ref{tab:run3perf} for the Run\,2 and Run\,3 data-taking conditions, respectively. The Deepset\,NN tagger consistently performs better than the combination of all classical taggers. Due to its inclusive nature the DeepSet\,NN tagger reaches a tagging efficiency of 100\% throughout. The tagging power of the DeepSet\,NN is about 20 to 25\% increased with respect to the combination of classical taggers for the Run\,2 samples and even more for the Run\,3 sample.

The tables also show an overall reduced tagging power on the Run\,3 with respect to the Run\,2 samples. This is due to the higher pileup in Run\,3, that leads to more particles in the event that are neither associated with the signal \B meson, nor do they carry information about its flavour. While the number of particles carrying information about the signal \B meson flavour stays the same between Run\,2 and Run\,3, the number of "background" particles increases significantly. In order to facilitate the DeepSet\,NN's task, we perform a selection on the input particles. Instead of using all charged particles in the event as inputs to the DeepSet\,NN, we select those that can be associated to the same primary vertex (PV)\footnote{The primary vertex is where the proton-proton interaction took place.} as the signal \B meson. Table~\ref{tab:run3perf} shows that this purification of the inputs leads to an increase in tagging power of $\sim$7\%.

\begin{table}[hb]
\caption{Comparison of the performance of the DeepSet\,NN tagger with the Classical taggers for the Run\,2 data-taking conditions of the \lhc and the \lhcb experiment. The tagging efficiency $\epsilon_\mathrm{tag}$ and the tagging power $\epsilon_\mathrm{eff}$ are shown for different signal \B meson decays.}
\centering
\label{tab:run2perf}  
\begin{tabular}{l|cc}
 & $\epsilon_\mathrm{tag}$[\%] & $\epsilon_\mathrm{eff}$[\%]  \\\hline
$\boldsymbol{\Bz \rightarrow \jpsi \Kstarz}$ & & \\
OS Combination & 38.5 & 3.81 \\
SS Combination & 80.1 & 1.71 \\
\textbf{Classical Taggers Combination} & 87.0 & 5.39\\
\textbf{DeetSet\,NN} & 100 & 6.38 \\
\hline
$\boldsymbol{\Bp \rightarrow \jpsi \Kp}$ & &\\
OS Combination & 38.2 & 3.94 \\
SS Kaon & 67.7 & 1.22 \\
SS Pion & 69.9 & 3.94 \\
\textbf{Classical Taggers Combination} & 92.0 & 6.39 \\
\textbf{DeepSet\,NN} & 100 & 8.0 \\
\hline
$\boldsymbol{\Bs \rightarrow \Ds \pim}$ & & \\
\textbf{DeepSet\,NN} & 100 & 8.7 \\
\end{tabular}
\end{table}

\begin{table}[hb]

\caption{Comparison of the performance of the DeepSet\,NN tagger with the Classical taggers for the Run\,3 data-taking conditions of the \lhc and the \lhcb experiment. The tagging efficiency $\epsilon_\mathrm{tag}$ and the tagging power $\epsilon_\mathrm{eff}$ are shown for $\Bp \rightarrow \jpsi \Kp$ meson decays.}
\label{tab:run3perf}  
\centering
\begin{tabular}{l|cc}
 & $\epsilon_\mathrm{tag}$[\%] & $\epsilon_\mathrm{eff}$[\%]  \\\hline
$\boldsymbol{\Bp \rightarrow \jpsi \Kp}$ & &\\
{Classical Taggers Combination} & 100 & 3.75 \\
{DeepSet\,NN} & 100 & 6.36 \\
{DeepSet\,NN same PV} & 100 & 6.83 \\
\end{tabular}
\end{table}

\section{Summary, Conclusion and Outlook}
\label{sec:summaryconclusionoutlook}
Determining the flavour of neutral \B mesons at production is essential for meson mixing and
time-dependent \CP violation measurements. The flavour tagging exploits information from particles that are created in correlation with the signal \B meson to obtain its flavour at production. The classical approach to flavour tagging is a set of individual algorithms that each look for only one specific particle that carries information about the flavour of the signal \B meson. These classical taggers suffer from a low tagging efficiency and require the training and evaluation of several different algorithms. In these proceedings, we propose an inclusive tagger which can consider the entire event as a whole. A suitable architecture for such an inclusive tagger is the DeepSet\,NN, which can take a list of inputs of variable length (since the number of input particles varies between events) and is invariant under the permutation of its inputs (since the ordering of the input particles should have no influence on the predicted flavour of the signal \B meson). In addition, the DeepSet\,NN architecture lends itself to parallelisation and is faster in training and evaluation than any previous flavour tagging algorithms. 
When compared on simulated data samples the flavour tagging performance of the DeepSet\,NN is consistently increased with respect to the combination of classical taggers. We also found that the tagging power of the DeetSet\,NN can be increased in the high-background environment of Run\,3 by performing a selection on the particles in the event prior to inputting them into the DeepSet\,NN.
In conclusion it can be said that the DeepSet\,NN shows very promising performances for the flavour tagging. It is also fast to train and to evaluate, which makes it suitable for future application in the real-time software trigger of the \lhcb experiment. The next steps are to evaluate and compare its performance on data and use it as part of a physics measurement.

\section*{Acknowledgements}
Niklas Nolte was supported by NSF grant PHY-2019786 (The NSF AI Institute for Artificial Intelligence and Fundamental Interactions, http://iaifi.org).
Claire Prouve was supported by Juan de la Cierva Incorporaci\'on - USC 2022 IJC2020-044329-I grant.


\begin{thebibliography}{}
\bibitem{Bsoscil}
LHCb collaboration, \href{https://doi.org/10.1038/s41567-021-01394-x}{Precise determination of the \Bs - \Bsb oscillation frequency}, Nat. Phys. 18, 1–5 (2022)
\bibitem{Bs2JpsiK}
LHCb collaboration, \href{https://link.springer.com/article/10.1140/epjc/s10052-019-7159-8}{Updated measurement of time-dependent CP-violating observables in $\Bs \rightarrow \jpsi \kaon \kaon$ decays}, Eur.Phys.J.C 79 (2019) 8, 706, Eur.Phys.J.C 80 (2020) 7, 601 (erratum)
\bibitem{sstaggers}
LHCb collaboration, \href{https://cds.cern.ch/record/1484021?ln=en}{Optimization and calibration of the same-side kaon tagging algorithm using hadronic \Bs decays in 2011 data}, Hadron Collider Physics Symposium 2012, Kyoto, Japan, 12 - 16 Nov 2012
\bibitem{ostaggers}
LHCb collaboration, \href{https://link.springer.com/article/10.1140/epjc/s10052-012-2022-1}{Opposite-side flavour tagging of \B mesons at the LHCb experiment}, Eur. Phys. J. C 72 (2012) 2022
\bibitem{RefDeepSetNN}
Manzil Zaheer, Satwik Kottur, Siamak Ravanbakhsh, Barnab{\'{a}}s P{\'{o}}czos, Ruslan Salakhutdinov and Alexander J. Smola, \href{http://arxiv.org/abs/1703.06114}{Deep Sets},  arXiv:1703.06114v3 [cs.LG]


\end{thebibliography}
\end{document}